\documentclass[usenatbib,letters]{mn2e}
\usepackage{amsmath}
\usepackage{amssymb}
\usepackage{graphicx}
\usepackage{aas_macros}

\author[B. Farris et al.]{Brian D.~Farris$^{1,2}$, Paul~Duffell$^{1}$, Andrew I.~MacFadyen$^{1}$, and Zolt\'an~Haiman$^{2}$ \\
    $^1$Center for Cosmology and Particle Physics, Physics Department, New York University, New York, NY 10003, USA\\
$^2$Department of Astronomy, Columbia University, 550 West 120th Street, New York, NY 10027, USA} 

\title{Binary Black Hole Accretion During Inspiral and Merger}

\begin{document}

\maketitle
\begin{abstract}
    We present the results of 2D, moving mesh, viscous hydrodynamical simulations of accretion onto merging supermassive black hole (SMBH) binaries. We include viscous heating, shock heating, and radiative cooling, and simulate the transition from the ``pre-decoupling" epoch, where the inspiral timescale is longer than the viscous timescale, to the ``post-decoupling" epoch, where the inspiral timescale is shorter than the viscous timescale. We find that there is no abrupt halt to the accretion at decoupling, but rather the accretion shows a slow decay, with significant accretion well after the expected decoupling. Moreover, we find that the luminosity in X-rays is significantly higher prior to the merger, as orbital energy from the SMBH binary is converted to heat via strong shocks inside the cavity, and radiated away. Following the merger, the cavity refills viscously and the accretion rate relaxes to the Shakura-Sunyaev value, while the X-ray luminosity drops as the shocks quickly dissipate. 
\end{abstract}

\section{Introduction}
Supermassive black holes are currently believed to reside in nearly all galactic nuclei \citep{kormendy13,ferrarese05}. When such galaxies merge, a combination of dynamical friction, gravitational slingshot interactions with stars, and interactions with gas can bring the black holes close enough to become gravitationally bound, forming a SMBH binary system in the merged galaxy remnant \citep{mayer13}. The nuclei of merged galaxy remnants are expected to contain an abundance of dense gas \citep{barnes92,springel05}, which may form a circumbinary accretion disk \citep{artymowicz96,armitage02,milos05}.

The standard picture for the evolution of the system is as follows: For large binary
separations $a$, the inspiral time due to gravitational wave (GW) emission is much longer
than the viscous time ($t_{\rm GW}\gtrsim t_{\rm vis}$), so that the disk
settles into a quasi-stationary state. For equal-mass BHs, the binary
tidal torques carve out a partial hollow in the disk
\citep{artymowicz94,milos05,macfadyen08,kocsis12b} of radius
$\sim 2a$ and excite spiral density waves throughout the disk, that
dissipate and heat the gas. However, gas can penetrate the hollow in
response to the time-varying tidal torque
\citep{macfadyen08,hayasaki08,farris11,kocsis12b,noble12,roedig12}.
At sufficiently small separations $t_{\rm GW} \lesssim t_{\rm vis}$,
and the BHBH \emph{decouples} from the disk as the viscosity is unable to refill the cavity quickly enough to follow the shrinking picture. As a result, it has been proposed that the final SMBH binary merger takes place in a vacuum, and that the associated EM signature can only appear after the cavity refills from the decoupling radius on a viscous timescale \citep{milos05,shapiro10,tanakamenou10}.

Such binaries may provide a unique opportunity to observe
electromagnetic signatures
originating from the interaction of the binary with the surrounding
accretion disk. The gravitational radiation originating from the SMBH
binary inspiral should also be detectable by Pulsar Timing Arrays
(PTAs) \citep{hobbs10,kocsis11,tanaka12,lommen12,sesana12} or by a space interferometer such
as eLISA \citep{amaroseoane13}, provided the binary has maintained a circumbinary disk past the decoupling epoch \citep{barausse08,noble12, kocsis12b}.

Recent simulations of such binaries in the pre-decoupling epoch have called this picture into question. Prior to decoupling, simulations have shown that the cavity can be significantly lopsided so that gas is able to approach the BHs much closer than $\sim 2a$. Moreover, the cavity is penetrated by accretion streams with surface densities comparable to that of the circumbinary disk itself, which allow the binary to accrete freely, with very little suppression due to the gravitational torques of the binary \citep{artymowicz94,macfadyen08,sesana12,shi12,noble12,dorazio13,farris14a}. In this paper, we simulate viscous accretion onto a binary, allowing the binary orbit to shrink due to gravitational radiation such that our simulations pass through the ``decoupling" epoch and proceed all the way to merger. We estimate the extent to which accretion is diminished due to decoupling. We also examine time-dependent signatures of the merger which may appear in the spectrum.

We consider equal-mass, nonspinning
binaries.  While the BH mass scales out, we are primarily interested
in total masses $M \approx 10^8 M_\odot$ and low density disks
for which the tidally induced inspiral is subdominant, and the disk
self-gravity is negligible.

\section{Methods}
Our initial disk configurations are similar to that of \citet{farris14b}, and consist of a steady-state Shakura-Sunyaev disk solution, modified by adding a hollow cavity within $r \lesssim r_0 \equiv 2.5 a$. We assume a geometrically thin, gas pressure dominated, optically thick accretion disk, with electron scattering as the dominant opacity. In the early stage of our simulations, the inner cavity wall migrates inward due to viscosity, accretion streams form, and a quasi-steady state is reached. The fluid evolves according the the 2D viscous Navier Stokes equations, assuming an $\alpha$-law viscosity prescription.

Following \citet{peters64}, we account for shrinkage of the binary separation $a$ due to gravitational radiation to leading order, 
\begin{equation}
    a = a_0 \left(1-\frac{t-t_0}{\tau}\right)^{1/4}
\end{equation}
where $a_0$ is the separation at initial separation at time $t_0$, and $\tau$ is the characteristic merger timescale. Throughout the inspiral, we assume that the binary remains on a circular orbit. Because $\tau$ depends on $a_0$, and our simulations can be rescaled to any separation, we need only choose $\tau$ large enough so that $\tau > t_{vis}$ so that the fluid near the binary can relax to a quasistationary configuration before ``decoupling" occurs and the shrinkage begins to outpace the viscous refilling of the cavity. Thus, we choose $\tau = 5 t_{vis}$.

\begin{figure}
    \centering  
    \includegraphics[width=0.99\columnwidth]{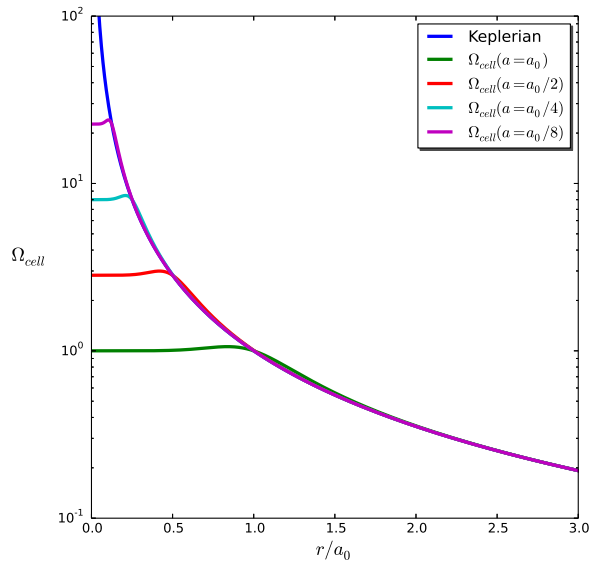}
    \caption{Profiles of the angular velocity of the computational cells at different binary separations. For $r>a(t)$, the motion is approximately Keplerian, and smoothly transitions to a constant angular velocity for $r<a$.}
    \label{fig:omega}
\end{figure}

The DISCO code \citep{duffell14} allows one the freedom to specify the motion of the computational cells. For this problem we choose a rotation profile which matches the nearly Keplerian fluid motion outside the cavity, while transitioning to uniform rotation at the binary orbital frequency inside the cavity. The radius at which this transition occurs shrinks with the binary according to
\begin{equation}
    \Omega_{cell} = \frac{r^{n-3/2}+a^{n-3/2}}{r^n+a^n} \ .
\end{equation}
where $n=8$. After the binary merges and $a=0$, all cells move at the Keplerian rate. Profiles of $\Omega_{cell}$ at various epochs are displayed in Fig.~\ref{fig:omega}.

For each simulation, we use a $\Gamma$-law equation of state of the form $P = (\Gamma - 1)\rho \epsilon$, where $P$ and $\epsilon$ are the z-integrated pressure and internal energy, respectively, and $\Gamma=5/3$ is the adiabatic index. Viscous heating  and radiative cooling  
are incorporated naturally through the energy equation, as described in \citet{farris14b}.
\section{Results}
\begin{figure*}
    \centering  
    \includegraphics[width=2.0\columnwidth]{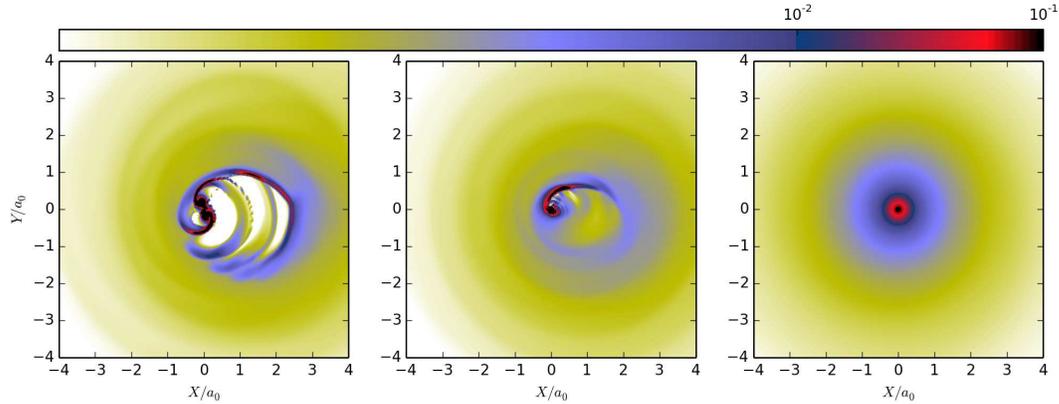}
    \caption{Snapshot of surface brightness dL/dA at (from left to right) $t-t_m = -0.1 t_{vis}, 0, 3 t_{vis}$, normalized by the value at r = a for a steady-state disk around a single BH.}
    \label{fig:emission}
\end{figure*}

\begin{figure}
    \centering  
    \includegraphics[width=0.99\columnwidth]{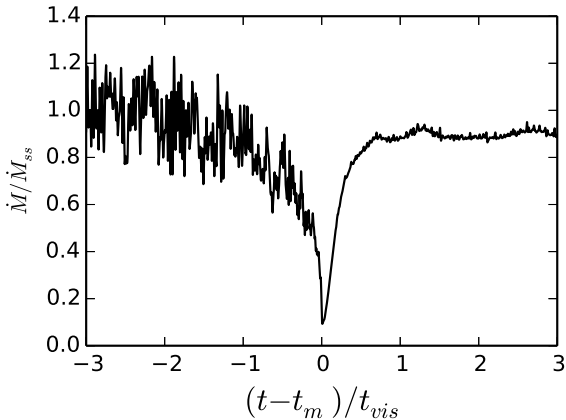}
    \caption{The time variable accretion rate onto the binary is plotted over a period of 6 viscous times. We note that the average accretion rate at early times is consistent with the expected Shakura-Sunyaev rate, and the post-merger refilling allows the circumstellar disk to relax to the approximate Shakura-Sunyaev solution for a single BH. }
    \label{fig:timeplot}
\end{figure}

Our initial disk contains an artificial cavity of radius $\approx 2.5 a$. Viscosity in the circumbinary disk transports angular momentum outward, causing the cavity to refill. Gravitational torques from the binary tend to drive matter outward, and a quasistationary balance between competing gravitational and viscous torques is achieved after $\sim 2 t_{vis}$. The quasisteady state is similar to that of \citet{farris14a,farris14b}, in that well collimated, narrow accretion streams form which penetrate the cavity, efficiently delivering fluid to the BHs. The cavity also becomes lopsided, as described in \citet{farris14a} and \citet{shi12}. One may naively think that when $t_{vis} = t_{gw}$, decoupling should occur and the accretion rate should drop essentially to zero. We find this not to be the case. Rather, we find a gradual decline in $\dot{M}$ beginning when $t_{vis} \approx t_{gw}$, but with significant accretion persisting until much closer to merger. This can be seen by examining the snapshot in Fig.~\ref{fig:emission}, in which accretion streams and minidisks are clearly present as late as $t-t_m = -0.1 t_{vis}$, where $t_m$ is the merger time, and in Fig.~\ref{fig:timeplot}, where we see that the accretion rate has only been reduced by a factor of $\approx 0.5$ by this time. We attribute the persistence of accretion at late times to the high-density accretion streams which are very effective at delivering gas into the cavity, and are not treated properly in axisymmetric 1D calculations.

\begin{figure}
    \centering  
    \includegraphics[width=0.99\columnwidth]{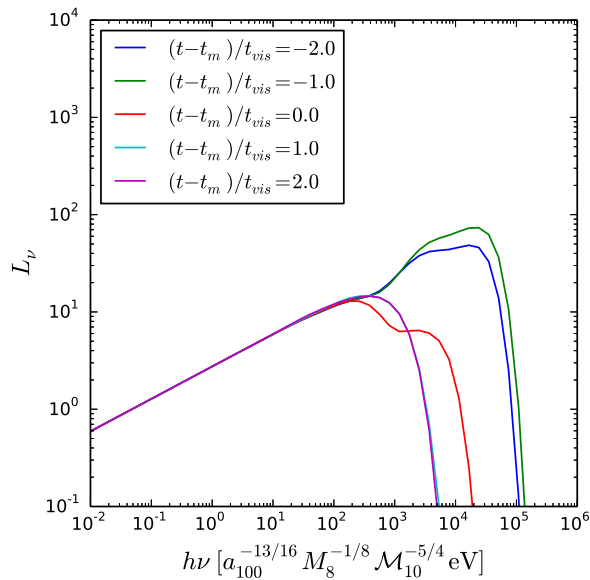}
    \caption{Emission spectrum at several epochs before, during, and after merger. High frequency emission prior to merger arises from dissipation of shock heating, and is absent following the merger.}
    \label{fig:spectrum}
\end{figure}

In order to examine changes in the spectrum throughout the inspiral and merger, we calculate the spectrum assuming thermal blackbody emission at surface patch in the disk,

\begin{equation}
    L_{\nu} = \int \frac{2 h \nu^3}{c^2\mbox{exp}\left(\frac{h \nu}{kT_{\rm{eff}}(r,\phi)}\right)-1}dA
\end{equation}
where the effective temperature $T_{\rm{eff}}(r,\phi) \equiv (q_{cool}(r,\phi) / \sigma)^{1/4}$, and $q_{cool}$ is the radiative cooling rate. In our simulations we allow heat generated through viscosity and shocks to be radiated at the blackbody rate instanteously, without accounting for the time necessary for photons to diffuse through the optically thick disk. As a result, some of the heat generated in the accretion streams may actually be retained and dissipated non-locally. The magnitude of this effect should be studied through future comparisons with 3D models which include radiative transfer. In Figure~\ref{fig:spectrum}, we plot spectra at $t/t_{vis} = -2.0, -1.0, 0.0, 1.0, 2.0$. We find that prior to merger, there is a strong enhancement in emission at high frequencies due to the shock heated minidisks. As in \cite{farris14b}, we find no noticeable ``notch" in the spectrum, as the low-frequency tail of the emission from the hot minidisks and the hot streams effectively washes out any defecits arising from the missing gas in the cavity.
Scaled to a $10^8 M_{\odot}$ binary  with a separation near decoupling at $a/M = 100$, we find that the enhancement is significant in soft and hard X-rays. Following the merger, there is a dramatic drop in high-frequency emission, as the shocks quickly dissipate in the absence of any binary torques.

In each curve in Fig.~\ref{fig:spectrum}, we note that there is a cutoff at high-frequencies. Physically, this corresponds roughly to the effective temperature at the BH horizon, which is the hottest region of the disk. We account for these horizons by masking out emission within a distance of $2M$ from each point mass, assuming that $a_0 = 100M$, which is of the order of the decoupling separation. This is admittedly a crude prescription, and we expect that the precise frequency at which the cutoff occurs will be modified in a fully relativistic simulation which accounts naturally for accretion through each BH's horizon. However, the qualitative result that this cutoff is at higher frequencies prior to merger, due to the shock heating of the minidisks and parts of the accretion streams, should be robust.

Comparing the spectra before and after merger, it is clear that the total bolometric luminosity is much greater prior to merger. This is somewhat counterintuitive, as circumbinary accretion disks have often been assumed in the literature to be dim as a result of binary torques choking off accretion. However, given the current understanding that accretion is not significantly diminished by binaries, the ability for gas in the cavity to tap into the large resevoir of orbital energy in the binary must be considered. As the precise amount of energy transferred from the binary depends on complicated 3 body interactions as well as the dynamics of the shock-heated gas in the cavity, a simple estimate for the magnitude of the bolometric luminosity enhancement prior to merger may not be possible. In our simulations, we find the enhancement to be roughly 2 orders of magnitude, but we caution that this result is very preliminary, and must be checked against simulations with more detailed treatments of dynamics of gas near the BH horizon. We also note that similar enhancements in bolometric luminosity due to tidal heating have been noted in previous analytic calculations \citep{kocsis12a,kocsis12b}.

\begin{figure}
    \centering  
    \includegraphics[width=0.99\columnwidth]{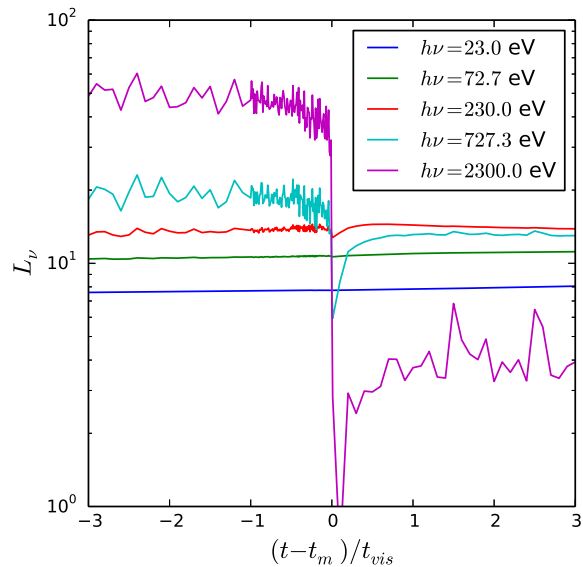}
    \caption{Lightcurves showing change in luminosity over the course of inspiral, merger and post-merger at a range of frequencies. Note that low frequency emission originates primarily from undisturbed circumbinary disk and exhibits little variation, while high frequency emission originates from shocked minidisks and streams and decreases significantly at merger.}
    \label{fig:lightcurves}
\end{figure}

\begin{figure}
    \centering  
    \includegraphics[width=0.99\columnwidth]{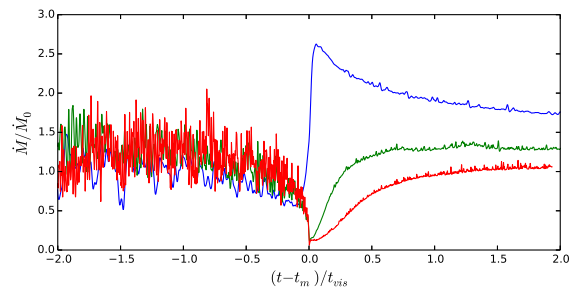}
    \caption{Evolution of accretion rate for three different values of $\mathcal{M}_a = 14 \mbox{ (red)}, 10 \mbox{ (green)}, 7 \mbox{ (blue)}$. In each case, time values are rescaled by the respective viscous timescale, and $\dot{M}$ is normalized by the respective value for a Shakura-Sunyaev disk around a single BH.}
    \label{fig:timeplots_PoRho}
\end{figure}

While the EM emission from an accreting binary can be modified significantly over the course of the inspiral and merger, past work has often focused on the total bolometric luminosity, ignoring the fact that lightcurves can be highly frequency dependent. Calculations which did include frequency evolution have assumed a lack of high-frequency emission from within the cavity and have focused on post-merger brightening (see e.g. \citealt{milos05,tanakamenou10}). In our simulations, we find that emission from the relatively cool gas in the circumbinary disk shows little change throughout the inspiral and merger, as the approximately Keplerian flows in this region are not significantly altered. As a result, the lightcurve at low frequencies shows very little variation. Emission from the shock heated gas in the minidisks and accretion streams, on the other hand, changes dramatically during the final stage of the inspiral and merger. As the shocks dissipate following merger, high frequency emission is dramatically reduced, and the lightcurves show sharp drops coincedent with the merger (see Fig.~\ref{fig:lightcurves}). This drop is opposite the brightening proposed previously \citep{milos05,tanakamenou10,shapiro10}.
We parametrize our simulations by the Mach number at $r=a_0$, $\mathcal{M}_a$. In order to probe potential dependence on $\mathcal{M}_a$, we have simulated mergers with $\mathcal{M}_a = 14,10,7$. In Fig.~\ref{fig:timeplots_PoRho}, we plot the accretion rate $\dot{M}$ for each case. Time values are rescaled by the respective viscous timescale, and $\dot{M}$ is normalized by the respective value for a Shakura-Sunyaev disk around a single BH. Prior to merger, the accretion rate is strikingly similar for each case, indicating that the gradual decoupling we observe is likely a generic effect. The rebrightening following merger does seem to vary significantly for the cases considered, which we attribute to the fact that the size of the cavity and its lopsidedness grows with $\mathcal{M}_a$, causing a longer delay in refilling for disks with large $\mathcal{M}_a$.

\section{Discussion}

Prior work on SMBH binaries has emphasized a ``post-decoupling" epoch, during which the binary separation shrinks due to gravitational radiation too quickly for the inner disk cavity to viscously refill, leading to mergers with very little gas present. We have performed 2D simulations of accretion onto a shrinking binary and shown that the ``decoupling" process is actually very gradual, and that significant accretion can persist well after the expected decoupling epoch, including separations corresponding to orbital frequencies that place the binary inside the PTA and LISA bands. This is encouraging as it suggests that EM counterparts for PTA and eLISA sources should be observable.

We have shown several observable electromagnetic signatures of SMBH mergers. We have computed blackbody spectra from snapshots of our simulations and shown that the high-frequency enhancements in the spectra which arise from shock heated minidisks and accretion streams disappear following the merger, as the shocks are no longer present in the absence of the binary torques. This effect is also reflected in steep drop in the high-frequency lightcurves coinciding with the binary merger. We note that the enhancement in UV/soft X-ray may also lead to stronger broad lines prior to merger.

In our simulations, we find the total enhancement in bolometric luminosity during the pre-merger phase to be roughly a factor of $\eta \sim 100$, so that $L \sim \eta G M \dot{M} / R_{isco}$. This extra energy is emitted by shock heated gas in the streams and minidisks, but ultimately comes from the orbital energy of the binary. In our calculations, we have neglected any changes to the binary orbit due to this effect, assuming it is small compared to the rate of energy loss due to gravitational waves, $L_{gw} \sim (GM/ac^2)^4 M c^3/a$. We can check our assumption by computing the ratio of these luminosities,
\begin{equation}
    \frac{L}{L_{gw}} \sim 10^{-3} 
    \left(\frac{\eta}{100}\right)
    \left(\frac{M_{bh}}{10^8 M_{\odot}}\right)
    \left(\frac{ac^2/GM}{100}\right)^5
    \left(\frac{\dot{M}c^2}{L_{edd}}\right) \ .
\end{equation}
Thus, we see that our neglect of energy loss from the binary due to interaction with the disk is warranted for our fiducial parameters. However, the above ratio is very sensitive to binary separation, and can easily exceed $\sim 1$ at moderate separations. In this case, not only will the inspiral rate be modified, but the binary eccentricity may grow due to interaction with the lopsided inner disk. This motivates the need for future calculations with a live binary, which self-consistently include the torques from the circumbinary disk and tracks the binary orbital evolution.

Our results our based on idealized 2D viscous hydrodynamic simulations with black-body cooling. In future work, we intend to make several improvements. These include adding radiation pressure in our calculations, as the shock-heated minidisks and accretion streams may become radiation pressure dominated. We also intend to relax our black-body assumption and take into account radiative processes such as inverse Compton scattering that may be relevant.  In this paper, we have focused our attention on equal-mass binaries, but we intend to study the dependence of spectral features on binary mass-ratio. We also intend to determine the dependence of these features on disk thicknesses. For computational efficiency, we have performed our simulations in 2D, assuming an $\alpha$-law viscosity as a proxy for the viscosity arising from magneto-hydrodynamic (MHD) turbulence. We also intend to perform full 3D MHD simulations using the {\it DISCO} code \citep{duffell14} in order to verify the validity of these approximations, and to replace our Newtonian prescription with a fully relativistic treatment so that gas dynamics near the horizons can be more precisely accounted for. We don't expect thise improvements to modify our basic conclusion that the binary is bright during the coalescence process, except for perhaps a brief pause at the merger itself, and this pause is much shorter than the fiducial viscous time. This is an encouraging finding for EM counterparts to GW sources. 

\section{Acknowledgements}
Resources supporting this work were provided by the NASA High-End Computing (HEC) Program through the NASA Advanced Supercomputing (NAS) Division at Ames Research Center and by the High Performance Computing resources at New York University Abu Dhabi. We acknowledge support from NASA grant NNX11AE05G (to ZH and AM). We are grateful to Daniel D'Orazio and Andrei Gruzinov for helpful comments and discussions. 

\bibliographystyle{mn2e}
\bibliography{refs}

\end{document}